\documentclass{aa}  
\usepackage{amsmath,scalerel}
\usepackage{graphicx}
\usepackage[colorlinks=true, linkcolor=blue, citecolor=blue, filecolor=blue, urlcolor=blue]{hyperref}
\usepackage{txfonts}
\usepackage[dvipsnames]{xcolor}
\usepackage{xcolor}
\usepackage{ulem}
\graphicspath{ {figures/} }
\begin{document}

   \title{Journey of complex organic molecules: Formation and transport in protoplanetary disks}
    \titlerunning{Journey of complex organic molecules: formation and transport in PPDs}
%    \subtitle{I. Overviewing the $\kappa$-mechanism}

   \author{T. Benest Couzinou\inst{1}, O. Mousis\inst{1,2}, G. Danger\inst{3,2}, A. Schneeberger\inst{1}, A. Aguichine\inst{4,1}, \and A. Bouquet \inst{3,1}, 
          }
    \authorrunning{Benest Couzinou et al.}

   \institute{
Aix-Marseille Université, CNRS, CNES, Institut Origines, LAM, Marseille, France
   \and
   Institut Universitaire de France (IUF), Paris, France
   \and
   Aix-Marseille Université, CNRS, Institut Origines, PIIM, Marseille, France
   \and
   Department of Astronomy and Astrophysics, University of California, Santa Cruz, CA, USA
   }

   \date{Accepted October 02, 2024}

% \: the case of methanol iceabstract{}{}{}{}{}
% 5 {} token are mandatory
 
  \abstract
   {Complex organic molecules serve as indicators of molecular diversity. Their detection on comets, planets, and moons has prompted inquiries into their origins, particularly the conditions conducive to their formation. One hypothesis suggests that the UV irradiation of icy grains in the protosolar nebula generates significant molecular complexity, a hypothesis supported by experiments on methanol ice irradiation.
   }
   {We investigated the irradiation of methanol ice particles as they migrate through the protosolar nebula. Our objective is to ascertain whether the encountered conditions facilitate the formation of complex organics molecules, and we leverage experimental data in our analysis.}
   {We developed a two-dimensional model that describes the transport of pebbles during the evolution of the protosolar nebula, employing a Lagrangian scheme. This model computes the interstellar UV flux received by the particles along their paths, which we compared with experimental values. }
   {On average, particles ranging from 1 to 100 $\mu$m in size, released at a local temperature of 20 K, undergo adequate irradiation to attain the same molecular diversity as methanol ice during the experiments within timescales of 25 kyr of protosolar nebula evolution. In contrast, 1 cm sized particles require 911 kyr of irradiation to reach similar molecular diversity, making comparable molecular complexity unlikely. 
   Similarly, particles ranging from 1 to 100 $\mu$m in size, released at a local temperature of 80 K, receive sufficient irradiation after 141 and 359 kyr. In contrast, 1 cm sized particles would require several million years to receive this level of irradiation, which is infeasible since they cross the iceline within approximately 500 kyr.
    }
  % conclusions heading (optional), leave it emspty if necessary
   {The particles readily receive the irradiation dose necessary to generate the molecular diversity observed in the experiments within the outer regions of the disk. Our model, combined with future irradiation experiments, can provide additional insights into the specific regions where the building blocks of planets form.}


    \maketitle

\section{Introduction} \label{sec:intro}

%Protoplanetary disks (PPDs) are the cosmic nurseries where dust grains grow into pebbles, planetesimals, comets, asteroids, as well as planets. Because comets and asteroids are less evolved than planets, the investigation of their composition provides hints about the thermodynamic conditions that were at play in the protosolar nebula (PSN), as well as about the time and radial evolution of its chemical and isotopic composition. Ices have been detected in molecular clouds and circumstellar disks \citep{van2004,Boo2015}, and their irradiation induces photochemistry leading to a molecular diversity through different energetic processes \citep{Dan2013,Obe2016}. Complex organic molecules (COMs) are molecules with several atoms of carbon, hydrogen and oxygen, such as \textrm{CH$_3$CH$_2$OH}, \textrm{CH$_3$OCH$_3$} or \textrm{C$_2$H$_5$OH} \citep{Bel2009,Ten2022}, and form markers of this diversity. Moreover, COMs have been found in meteorites and comets, and are observed in star-forming regions \citep{Bri1992,Bis2007,Ber2017}.

Protoplanetary disks (PPDs) serve as cosmic nurseries where dust grains evolve into pebbles, planetesimals, comets, asteroids, and eventually planets. Given that comets and asteroids are less evolved than planets, analyzing their composition offers insights into the thermodynamic conditions present in the protosolar nebula (PSN) and the temporal and radial evolution of its chemical and isotopic composition. Ices have been identified in molecular clouds and circumstellar disks \citep{van2004,Boo2015} undergoing photochemistry-induced molecular diversity via various energetic processes \citep{Dan2013,Obe2016}. Complex organic molecules (COMs) are molecules composed of multiple atoms of carbon, hydrogen, and oxygen, such as \textrm{CH$_3$CH$_2$OH}, \textrm{CH$_3$OCH$_3$}, or \textrm{C$_2$H$_5$OH} \citep{Bel2009,Ten2022}, and have been discovered in meteorites, comets, and star-forming regions \citep{Bri1992,Bis2007,Ber2017}.

Laboratory experiments have demonstrated that COMs form through the alteration of astrophysical ices via UV irradiation, cosmic ray bombardment \citep{Ehr2000,She2004}, and electron \citep{Mai2015} or ion \citep{Moo1991,Hud1999} bombardment in PSN conditions. Among these ices, methanol is particularly abundant \citep{Pon2003,Gib2004,van2014} and serves as an ideal molecule for generating significant molecular diversity \citep{Obe2009,Cup2011}: Lyman-$\alpha$ irradiation of pure methanol ice at 20 K results in the detection of over 35 gas-phase molecules with 2 to 6 carbon atoms \citep{Abo2014}. The generation of this diverse array of COMs primarily depends on the dose received by the ice and the temperature at which the ice undergoes processing \citep{The2013, Ten2022}. 

While this study primarily investigates ice processes within the PSN, it is important to note that alternative experiments suggest COMs could have also formed in the interstellar medium (ISM), where the local temperature is $\sim$10 K in dark molecular clouds \citep{Com1995,Qua2020,Car2021,Fed2022}. However, due to the complexity of PPDs involving various physical processes (turbulence, diffusion, advection, UV irradiation, etc.), the formation conditions of COMs have not been extensively explored in these environments. Irradiation levels notably fluctuate across the disk, ranging from higher rates at the surface to near absence in the midplane region \citep{CieSan2012,Mou2018}. Consequently, the trajectory of icy particles in PPDs can significantly influence their photochemical evolution.

In this study we computed the irradiation dose received by grains and icy pebbles during their radial and vertical transport using a time-dependent PPD model derived from \cite{Agu2020} and \cite{Sch2023}, following a Lagrangian scheme \citep{Cie2010,Cie2011,Ron2017}. The results were then compared with experimental data derived by \cite{Ten2022}, who irradiated methanol ices with a UV photon flux in a vacuum chamber to track COM formation and composition. This comparison allowed us to evaluate whether the formation conditions of COMs from methanol ice identified in the laboratory could occur in certain regions of the PSN.

Section \ref{sec:experiments} describes the laboratory experiments used to benchmark our transport and irradiation model. Section \ref{sec:method} details the different modules of our model, while Sect. \ref{sec:results} presents the comparisons between our numerical results and the experiments. Finally, Sect. \ref{sec:discussion} is dedicated to a discussion and conclusions.

\section{Experimental frame} \label{sec:experiments}

The experimental data utilized in our paper were derived from the research conducted by \cite{Ten2022}. A pure methanol ice underwent irradiation at Lyman-$\alpha$ under varying doses with a lamp flux of $10^{13}$ photons cm$^{-2}$ s$^{-1}$ \citep{Ger1996}. Irradiation times of up to 24 hours were tested at both 20 and 80 K, resulting in an accumulation of up to $8.64 \times 10^{17}$ photons/cm$^2$ on the icy sample. The icy sample consists of a thin layer of methanol, containing approximately $1.2 \times 10^{17}$ molecules per cm$^2$, resulting in a photon--to--molecule ratio of 7.2. After irradiation, the sample was returned to room temperature, and the volatiles were released into the gas phase. Volatile organic compounds were then characterized using a gas chromatograph coupled with a mass spectrometer. Twenty-one different organic molecules containing up to five carbon atoms (including alcohols, aldehydes, ketones, esters, and ethers) were identified. The results demonstrate that molecular diversity increases with the irradiation dose from only MeOH. Interestingly, the abundances of C$_1$ and C$_2$ molecules tend to decrease or stabilize with increasing irradiation dose, while heavier molecules emerge. However, lower abundances of heavier molecules are present, consistent with radical reactivity. Among all the chemical families observed, ketones and esters contain the molecules with the largest number of carbon atoms.

Experiments conducted at 80 K require a higher irradiation dose to achieve the same molecular complexity observed at 20 K. In both cases, the data analyses indicate that 24 hours of irradiation suffice to enable COM formation at both temperatures, corresponding to a dose of $8.64 \times 10^{17}$ photons cm$^{-2}$.

Our modeling aims to ascertain the conditions under which COMs form from irradiated methanol icy pebbles or grains during their migration in the PSN. To achieve this, CH$_3$OH-rich particles are released in PSN regions where the temperature is initially set to 20 or 80 K, the two temperatures considered in the experimental setup. Irradiation accumulates along the particles' drift until it reaches the necessary doses for COM formation ($8.64 \times 10^{17}$ photons.cm$^{-2}$). Successful particles are those that reach the irradiation dose identified in experiments before crossing the methanol iceline, at approximately 124 K. Once the particles cross this iceline, they are assumed to sublimate.

\section{Methodology} 
\label{sec:method}

This section describes the modules employed for computing the evolution of the disk, as well as the transport and irradiation of the particles.    

\subsection{Structure of the disk} 
\label{sec:diskevol}
    
Our time-dependent accretion disk model is computed in the framework of the $\alpha$-turbulent viscosity formalism, in which the viscosity is defined as $\nu = \alpha c_{\mathrm{s}}^2 / \Omega_{\mathrm{K}}$ \citep{Sha1973}. Here, $\Omega_{\mathrm{K}} = \sqrt[]{GM_{\mathrm{\star}} / r^3}$ is the Keplerian frequency, $M_{\mathrm{\star}}$ is the mass of the star, $r$ is the distance to the star, and $G$ is the gravitational constant. $c_{\mathrm{s}} = \sqrt[]{R_{\mathrm{g}} T_{\mathrm{m}} / \mu}$ is the isothermal sound speed, $R_{\mathrm{g}}$ is the ideal gas constant, $T_{\mathrm{m}}$ is the midplane temperature of the disk, and $\mu = 2.31$ g.mol$^{-1}$ is the mean molar mass of the gas. $\alpha$ is the viscosity parameter, usually set in the $10^{-4}$--$10^{-2}$ range \citep{Har1998,Hue2005,Des2017}, which we set to $10^{-3}$ in our model.
    
Our model computes the evolution of the gas surface density, $\Sigma_{\mathrm{g}}$, at hydrostatic equilibrium via the following differential equation \citep{Lyn1974},
        
\begin{equation} \label{eq:dSigmaG/dz}
\frac{\partial \Sigma_{\mathrm{g}}}{\partial t} = \frac{3}{r} \frac{\partial }{\partial r} \left [r^{ \frac{1}{2} } \frac{\partial }{\partial r} \left ( r^{ \frac{1}{2} } \Sigma_{\mathrm{g}} \nu \right ) \right ],
\end{equation}

\noindent which can be rewritten as

\begin{equation}
\begin{split}
\frac{\partial \Sigma_\mathrm{g}}{\partial t} = \frac{1}{2 \pi r} \frac{\partial \dot{M} }{\partial r} \quad \quad \\
\quad \quad \dot{M} = 3 \pi \Sigma_\mathrm{g} \nu \left ( 1+2 \frac{ \textup{dln}( \Sigma_\mathrm{g} \nu)}{\textup{dln}r} \right ),
\end{split}
\end{equation}

\noindent where $\dot{M}$ is the mass accretion rate.%, \tb{\sout{ which can be written as a function of the gas velocity field : $\dot{M} = -2 \pi r \Sigma_\mathrm{g} v_\mathrm{g}$}}.

Our code resolves this equation iteratively, using the following initial conditions \citep{Lyn1974,Agu2020}:
\begin{equation}
\left \{
\begin{array}{ll}
\Sigma_{\mathrm{g},0} = \frac{\dot{M}_{\mathrm{acc},0}}{3 \pi \nu} e^{-\left (\frac{r}{r_\mathrm{c}} \right )^{0.5} }
\\
\dot{M}_0 = \dot{M}_{\mathrm{acc},0} \left ( 1 - \left ( \frac{r}{r_\mathrm{c}} \right )^{0.5} \right ) e^{-\left (\frac{r}{r_\mathrm{c}} \right )^{0.5} }
\end{array}
\right.
,\end{equation}
\noindent where $r_\mathrm{c}$ is the centrifugal radius (computed here as 1.2 AU at initialization), and $\dot{M}_{\mathrm{acc},0}$, the accretion rate, is assumed to be $10^{-7.6} M_\odot\, \text{yr}^{-1}$.

%\noindent where $r_\mathrm{c}$ is the centrifugal radius \tb{(computed here as 1.2 AU)}, and assuming $\dot{M}_{\mathrm{acc},0} = 10^{-7.6} M_\odot .yr^{-1}$.
        
The temperature in the midplane is given by the sum of all the heating sources \citep{Hue2005}:

\begin{equation} 
\label{eq:temperature}
\begin{split}
T_\mathrm{m}^4 & = \frac{1}{2 \sigma_\mathrm{sb}} \left ( \frac{3}{8} \tau_\textup{R} + \frac{1}{2 \tau_\textup{P}} \right ) \Sigma_\mathrm{g} \nu \Omega_\mathrm{K}^2  + T_\mathrm{amb}^4,
\end{split}
\end{equation}

\noindent where $\sigma_\mathrm{sb}$ is the Boltzmann constant. $\tau _\textup{R} = \frac{\kappa_\textup{R} \Sigma_\mathrm{g}}{2}$ and $\tau _\textup{P} = 2.4 \tau _\textup{R}$ are the Rosseland and Planck optical depth, respectively. The mean Rosseland opacity, $\kappa_\textup{R} = \kappa_0 \rho_\mathrm{m}^a T^b$, is a function of the gas density and temperature in the midplane, with $\kappa_0$, $a$, and $b$ determined by observations (see \citealt{Bel1994} and our Appendix \ref{appendix:opacity}). Irradiation from the young star is neglected due to the assumption of shadowing in the outer part of the disk \citep{Ohn2021}. This shadowing is attributed to dust pileup at the water snowline and other icelines, which are inherent in our model \citep{Agu2020, Sch2023}. Finally, $T_\mathrm{amb} = 10~K$ is the ISM temperature in dark molecular clouds.

Assuming the disk is vertically isothermal (a common assumption, as the adiabatic disk is isothermal except in the inner regions; \citealt{Cie2010, Ron2017}), its vertical density, temperature, and pressure profiles are derived as follows:
    
\begin{equation} \label{eq:vertical_structure}
\rho_{\mathrm{g}}(r,z) = \rho_{\mathrm{m}}(r) e^{-\frac{z^2}{2H^2}}, \quad T(z,r) = T_{\mathrm{m}}(r), \quad P(z,r) = P_{\mathrm{m}}(r) e^{-\frac{z^2}{2H^2}},
\end{equation}
    
\noindent with $\rho_{\mathrm{m}}$, $T_{\mathrm{m}}$, and $P_{\mathrm{m}}$ corresponding to the midplane density, temperature, and pressure respectively, and $z$ is the vertical position. From those quantities, we define the gas disk scale height $H = c_{\mathrm{s}} / \Omega_{\mathrm{K}}$. We refer the reader to \cite{Agu2020} for a full description of our disk model. 

The temperature and pressure at which methanol condenses in the disk is computed using the Antoine's law:

\begin{equation}
\log_{10} (P_{\mathrm{vap}}) = A - \frac{B}{T+C}
\end{equation}

\noindent where A = 5.20409, B = 1581.341, C = -33.5, as provided by \cite{Amb1970} and obtained from the NIST Chemistry Webbook\footnote{\citeauthor{NIST}, https://webbook.nist.gov/chemistry/}, and $P_{\mathrm{vap}}$ is the vapor pressure in bar. The methanol iceline is defined as the location in the disk where the partial pressure of methanol is equal to its equilibrium pressure. The methanol gas phase abundance assumed in the disk is $1.4018~\times~10^{-6}$ with respect to H$_2$, based on an estimate of the CH$_3$OH/H$_2$O ratio in Comet 67P/Churyumov-Gerasimenko \citep{LeR2015}, and on a plausible water abundance with respect to H$_2$ in the PSN \citep{Sch2023}. 

\subsection{Irradiation of particles} 
\label{sec:irradiation}
    
The irradiation of the disk was computed via the following relation for a flux $F(r,z)$ of UV photons \citep{Cie2010,CieSan2012}:
    
\begin{equation}
F(r,z) = F_0 e^{-\tau (r,z)}
\end{equation}
    
\noindent with
    
\begin{equation} 
\tau (r,z) = \int_{|z|}^{\infty} \rho_{\mathrm{g}} (r,z) \kappa \mathrm{d}z,
\end{equation}
    
\noindent where $\tau (r,z)$ represents the optical depth of suspended matter above the particle, calculated at a given altitude and distance from the star, and $\kappa$ denotes the opacity, which is assumed to be equal to the frequency averaged mean Rosseland opacity derived from experimental data \citep{Bel1994,Agu2020,Sch2023}. $F_0$ is the incident UV interstellar flux equal to $10^8$ photons cm$^{-2}$ s$^{-1}$ (= 1 G$_0$). 1 $G_0$ (in Habing units) is equal to $10^{12}$ photons m$^{-2}$ s$^{-1}$, and corresponds to the quantity characterizing the average interstellar radiation field with an energy $h\nu$ bracketed between 6 and 13.6 eV (between 91.2 and 206.7 nm). Here, $\nu$ is the frequency and $h$ is the Planck constant. $G_0$ can be expressed as follows \citep{Yeg2009}:
    
\begin{equation}
G_0 = \frac{4 \pi \displaystyle{\int _{\nu (206.7 nm)} ^{\nu(91.2 nm)}} J_{\nu} {\mathrm{d}}\nu }{1.6 \cdot 10^{-3}}
,\end{equation}
    
\noindent with J$_\nu$ the intensity of the radiation in erg cm$^{-2}$ $s^{-1}$ sr$^{-1}$ Hz$^{-1}$. We used $G_0 = 9.986 \cdot 10^{8}$ eV cm$^{-2}$ s$^{-1}$ for the UV interstellar (isotropic) photon flux \citep{Hab1968,Par2000,Yeg2009,CieSan2012}. However, depending on the source, the incident interstellar source can be a factor of 10 higher if it comes from the spiral arms of a galaxy \citep{Yeg2009}, and from 10 to 3$\times$10$^4$ higher if it comes from clusters, where the median flux value experienced by a forming stellar system is 900 G$_0$ \citep{Ada2006}. We assumed in our calculations that the incident flux $F_0$ arrives perpendicularly to the disk, with the same intensity at each point of its surface, and that the irradiation from the central star is neglected \citep{Ohn2021}.
    
\subsection{Vertical and radial transport of particles} \label{sec:transport}

For this study, we used particles ranging in size from 1 $\mu$m to 1 cm, with a density of 1 g/cm$^3$. We assumed that the simulated particles do not collide with each other and that they remain constant in size. Given their size and density, the particles evolve according to the Epstein regime. In the PSN environment, particles are subject to specific drag regimes: the Stokes regime for larger particles and the Epstein regime for smaller particles. Particles are in the Epstein regime if they are smaller than the mean free path of the gas, $\lambda = \frac{m_{H_2}}{\pi \rho_g r^2_{H_2}} $, times 2.25 \citep{Ste1996, Sup2000}, where $m_{H_2}$ and $r_{H_2}$ are the mass and radius of H$_2$ molecules, and $\rho_g$ is the density of the H$_2$ gas. In our model, particles up to 1 cm in size are in the Epstein regime when they are located beyond 0.5 AU from the star. Furthermore, considering their Stokes number, only the larger particles, approximately 1 cm in diameter, are subject to radial drift, while the smaller particles remain predominantly coupled to the gas. This is supported by the particles' Stokes numbers, which range from $10^{-2}$ to $10^{-7}$ (see Appendix \ref{appendix:regimes}).

\subsubsection{Vertical transport} \label{sec:vertical}

The vertical transport of particles is depicted following a Lagrangian approach \citep{Cie2010}. Their vertical evolution is depicted via the following advection diffusion equation \citep{Dub1995, Gai2001,Cie2010}:
    
\begin{equation} \label{eq:adv,diff}
\frac{\partial \rho_i}{\partial t} = \frac{\partial}{\partial z} \left( \rho_\mathrm{g} D \frac{\partial \frac{\rho_i}{\rho_\mathrm{g}}}{\partial z} \right) - \frac{\partial}{\partial z}\left( \rho_i v_z \right),  
\end{equation}
    
\noindent where $\rho_i$ is the material density, $\rho_\mathrm{g}$ the gas density, $D = \frac{\nu}{1+\mathit{St}^2}$ the diffusivity, $\mathit{St}$ the Stokes number, and $v_z$ the dust vertical velocity due to gravitational forces.
%We developed it :
%\begin{equation}
%\frac{\partial \rho_i}{\partial t} = \frac{\partial^2}{\partial z^2} \left( D \rho_i \right) - \frac{\partial }{\partial z}\left[ \left( \frac{D}{\rho_{\mathrm{g}}}\frac{\partial \rho_{\mathrm{g}} }{\partial z} + v_z \right) \rho_i \right]
%\end{equation}

The vertical motion of an individual particle is computed using a Monte Carlo approach considering the random motion due to the viscous effects from eddies in the turbulent fluid \citep{Cie2010}: 

\begin{equation}
z_i = z_{i-1} + v_{\mathrm{adv},z} \delta t + R_1 \left [ \frac{2}{\sigma^2}D(z)\delta t \right ]^{\frac{1}{2}}
,\end{equation}

\noindent where $z_{i-1}$ is the initial vertical position of the particle. $z_{i}$ is its position after a given time step $\delta t = \frac{1}{\Omega_K}$. $R_1 \in [-1;1] $ is a random number, and $\sigma^2$ is its distribution variance (= 1/3 for uniform distribution) representing the random movement due to turbulence \citep{Vis1997}. Hence, two particles beginning their trajectories at the same location will move along distinct paths.

The advection velocity term has three components coming from the advective term analogy of the advection diffusion equation \citep{Cie2010,Ron2017,Mou2018}:
    
\begin{equation} \label{eq:v_adv,z}
v_{\mathrm{adv},z}=v_{z} + v_{\mathrm{gas},z} + \frac{\partial D}{\partial z}.
\end{equation}
    
\noindent 
% The first term of Eq. \ref{eq:v_adv,z} is calculated in the frame of the Epstein regime for small particles, where i) the relative velocity between particles and gas $v_{rel}$ is much smaller than the gas thermal sound velocity $v_{th}$, and ii) the particles' size is smaller than the mean free path of the gas $\lambda$ \citep{Bai1965,Ste1996,Sup2000,Per2011}. 
The $v_z$ is then given by the terminal velocities of the particles, with the drag forces taken into account \citep{Cuz2006,Cie2010}:
% \citep{Dub1995,Gar2003,Per2011,Ron2017}
    
\begin{equation} \label{eq:dv_z}
v_z = -t_s \Omega_K^2 z 
.\end{equation}
    
\noindent Here, $t_\mathrm{s}$ is the stopping time, which is the timescale required for the dust, assumed to consist of spherical grains, to transfer all its angular momentum to the gas. In the Epstein regime, $t_\mathrm{s}$ is given by \citep{Per2011,Mou2018}

\begin{equation} \label{eq:t_s}
%\begin{split}
t_\mathrm{s} = \frac{ \rho_\mathrm{s} }{\rho_\mathrm{g} } \frac{R_\mathrm{s}}{v_\mathrm{th}}
 %& = \left \{
    %\begin{array}{ll}
    %\scaleto{\frac{ \rho_\mathrm{s} }{\rho_\mathrm{g} } \frac{R_\mathrm{s}}{v_\mathrm{th}}}{20pt} \quad \quad \quad \quad \text{Epstein regime}
    %\\
    % \scaleto{\frac{8}{3} \frac{ \rho_\mathrm{s} }{\rho_\mathrm{g} } \frac{R_\mathrm{s}}{v_\mathrm{rel}} \frac{1}{C_D(Re)}}{20pt} \quad \text{Stokes regime}
    %\end{array}
 %   \right. \\
 %& = \left ( \frac{ \rho_\mathrm{g} v_\mathrm{th}}{\rho_\mathrm{s} R_\mathrm{s}} \mathrm{min} %\left [ 1, \frac{3}{8} \frac{v_\mathrm{rel}}{v_\mathrm{th}} C_D(Re) \right ] \right )^{-1}
%end{split}
,\end{equation}

\noindent with $v_\mathrm{th} = \sqrt[]{8/\pi} c_\mathrm{s} $ the thermal sound velocity, $R_s$ the solid dust grain radius and $\rho_\mathrm{s}$ its density (here 1 g.cm$^{-3}$).

The second term $v_{\mathrm{gas},z}$ of Eq. \ref{eq:v_adv,z} is derived from the advection term of Eq. \ref{eq:adv,diff}, and comes from the vertical variation of density of nebular gas, which is the medium through which the species is diffusing \citep{Cie2010,Ron2017,Mou2018}:

\begin{equation} \label{eq:v_gaz,z}
v_{\mathrm{gas},z} = \frac{D}{\rho_\mathrm{g}} \frac{\partial \rho_\mathrm{g}}{\partial z}
.\end{equation}
    
\noindent Combining Eqs. \ref{eq:vertical_structure} and \ref{eq:v_gaz,z}, we derive $v_{\mathrm{gas},z} = -D \frac{z}{H^2}$. 

The final term in Eq. \ref{eq:v_adv,z} considers the vertical variation of the diffusion coefficient, with its value assumed to be zero under the assumption of a vertically isothermal disk.

\subsubsection{Radial transport} \label{sec:radial}

A similar method, based on the one-dimensional radial advection-diffusion equation, was used to compute the radial transport of particles \citep{Gai2001,Cie2011}:

\begin{equation}
\frac{\partial \rho_i}{\partial t} = \frac{1}{r}\frac{\partial}{\partial r} \left( r \rho_\mathrm{g} D \frac{\partial \frac{\rho_i}{\rho_\mathrm{g}}}{\partial r} \right) - \frac{1}{r}\frac{\partial }{\partial r}\left( r \rho_i v_\mathrm{r} \right).
\end{equation}

\noindent In Cartesian coordinates, the equation becomes \citep{Cie2011}

\begin{equation}
\frac{\partial \rho_i}{\partial t} = \frac{\partial}{\partial x} \left( \rho_\mathrm{g} D \frac{\partial \frac{\rho_i}{\rho_\mathrm{g}}}{\partial x} \right) - \frac{\partial }{\partial x}\left( \rho_i v_x \right) + \frac{\partial}{\partial y} \left( \rho_\mathrm{g} D \frac{\partial \frac{\rho_i}{\rho_\mathrm{g}}}{\partial y} \right) - \frac{\partial }{\partial y}\left( \rho_i v_y \right)
,\end{equation}

\noindent where $v_x$ and $v_y$ are the components of $v_r$. This allows the description of the evolution of particles along the $x$ and $y$ axes in a way similar as the one provided for their motion along the $z$ axis \citep{Cie2011}:

\begin{equation}
\begin{array}{ll}
 x_i = x_{i-1} + v_{\mathrm{adv},x} \delta t + R_1 \left [ \frac{2}{\sigma^2}D(x)\delta t \right ]^{\frac{1}{2}}
\\
 y_i = y_{i-1} + v_{\mathrm{adv},y} \delta t + R_1 \left [ \frac{2}{\sigma^2}D(y)\delta t \right ]^{\frac{1}{2}}
\end{array}
,\end{equation}

\noindent where $x_{i-1}$ (resp. $y_{i-1}$) is the initial horizontal $x$ (resp. $y$) position of the particle. $x_{i}$ (resp. $y_{i}$) is its horizontal $x$ (resp. $y$) position after a given time step $\delta t = \frac{1}{\Omega_K}$. $R_1 \in [-1;1] $ is a random number, and $\sigma^2$ is its distribution variance (= 1/3 for uniform distribution) representing the random movement due to turbulence \citep{Vis1997}.

The advection velocity along the $x$ axis becomes the sum of the three terms:

\begin{equation} \label{eq:v_adv,x}
v_{\mathrm{adv},x} = v_{x} + v_{\mathrm{gas},x} + \frac{\partial D}{\partial r}\frac{x_{i-1}}{r_{i-1}}
.\end{equation}

\noindent The first term is the velocity of the material due to gas drag:

\begin{equation}
v_{x} \left( x_{i-1}, y_{i-1} \right) = v_r \frac{x_{i-1}}{r_{i-1}},
\end{equation}
\noindent where

\begin{equation} \label{eq:v_r}
v_r = -\frac{ 1}{1 + \mathrm{St}^2}\frac{3}{2} \frac{\nu}{r} + \frac{2\mathrm{St} }{1+\mathrm{St}^2}\frac{c_s^2}{r\Omega_K} \frac{\mathrm{dln}P}{\mathrm{dln}r}.
%1/(1+St²) * v_g + 2St/(1+St²)*v_drift
\end{equation}

\noindent The first term of $v_r$ represents the radial velocity due to the gas velocity for a steady disk with a mass profile in $r^{-1}$ \citep{Lyn1974,Cla1988,Gai2001,Cuz2003,Cie2011,Gui2014}, while the second term accounts for the radial drift \citep{Wed1977,Nak1986,Bir2012,Agu2020}.

The second term in Eq.\ref{eq:v_adv,x} is the same as its equivalent along the z axis and accounts for the density gradient in the midplane:

\begin{equation}
v_{\mathrm{gas},x} = \frac{D}{\rho_\mathrm{g}} \frac{\partial \rho_\mathrm{g}}{\partial r} \frac{x_{i-1}}{r_{i-1}}.
\end{equation}

The third term of Eq. \ref{eq:v_adv,x}, $\frac{\partial D}{\partial t}\frac{x_{i-1}}{r_{i-1}}$, is identical to the last term of Eq. \ref{eq:v_adv,z}, and corresponds to the contribution due to the diffusivity gradient along the x-axis. Equations along the y-axis can be developed in the same way as for the x-axis since axial symmetry is assumed.

\section{Results}  \label{sec:results}

We conducted our computations with the goal of replicating the experimental conditions outlined in Sect. \ref{sec:experiments}. This methodology enabled us to explore the conditions under which COMs form within the PSN. Each particle trajectory presented was averaged over a total of 500 independent particles released in the midplane under identical conditions, ensuring a statistically significant sample.

\subsection{Map of disk irradiation}  \label{sec:cartography}

\begin{figure}[t]
\centering
\includegraphics[width=\columnwidth]{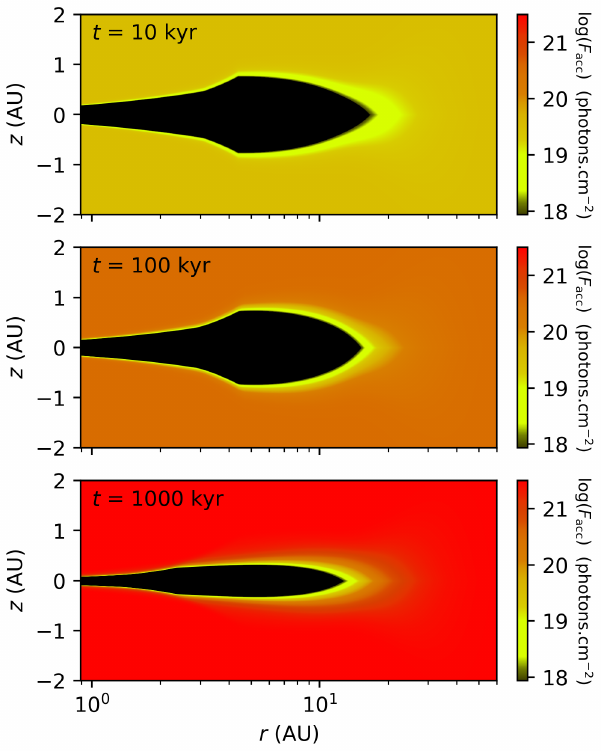}
\caption{Two-dimensional map of the accumulated irradiation after 10, 100, and 1000 kyr of the PSN's evolution. The black area around the midplane corresponds to the region where the accumulated irradiation is lower than the most important dose obtained during the experiments of interest ($8.64 \times 10^{17}$ photons.cm$^{-2}$).}
\label{fig:Fint(t)}
\end{figure}

Figure \ref{fig:Fint(t)} shows a two-dimensional map of $F_{\mathrm{acc}}$, namely the UV irradiation accumulated by the disk over its time evolution. As the disk's density shrinks and flattens over time, radiation progressively penetrates deeper into the midplane. After 10 kyr of disk evolution, the black zone in the figure, where the accumulated irradiation is lower than the most important dose obtained during the experiments of interest ($8.64 \times 10^{17}$ photons.cm$^{-2}$), extends radially up to $\sim$16.5 AU, and up to a height of $\sim$0.75 AU. After 100 kyr of irradiation, it reduces radially down to $\sim$15.1 AU, and slightly vertically down to $\sim$0.73 AU. After 1000 kyr of disk evolution, the black zone continues its shrinking radially down to $\sim$12.7 AU, and vertically down to 0.31 AU. At any epoch of the disk evolution, its outer regions reach accumulated irradiation levels exceeding by far the most important dose obtained during the experiments.

\subsection{Drift of irradiated particles} \label{sec:drift}

\begin{figure*}[h]
\centering
\includegraphics[width=\textwidth]{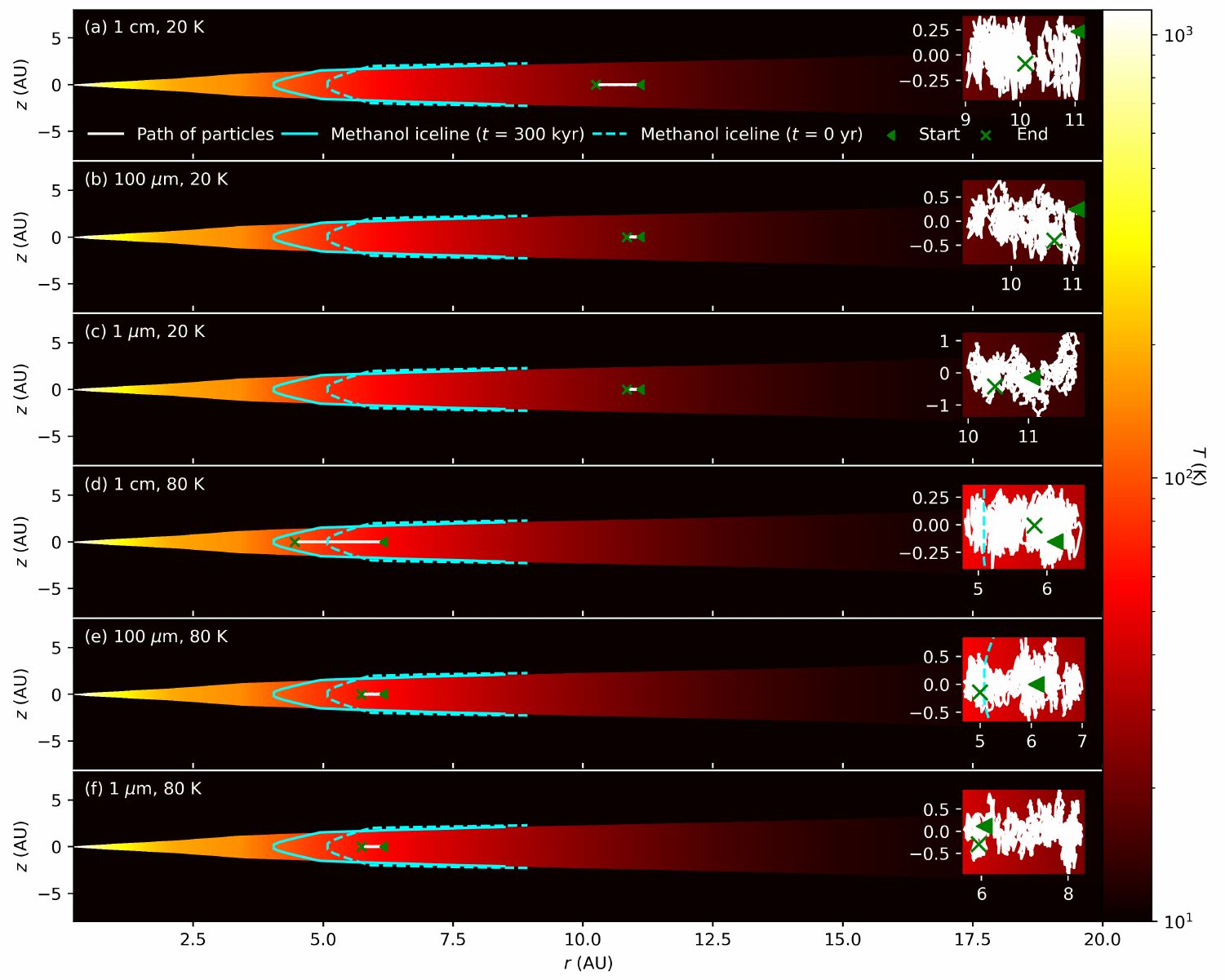}
\caption{Two-dimensional temperature map of the disk after 300 kyr of evolution, showing the methanol iceline at $t$ = 0 (dashed blue line) and at $t$ = 300 kyr (solid blue line). The white curve represents the average trajectory of 500 simulated particles. To provide a clearer illustration of individual particle behavior, the insets on the right display the trajectory of a single particle within the disk. The panels illustrate the trajectories of particles with sizes of 1 cm, 100 $\mu$m, and 1 $\mu$m, released at disk temperatures of 20 K (panels a, b, and c) and 80 K (panels d, e, and f). Initial and final positions are marked by a green triangle and a green cross, respectively.}
\label{fig:2d_map}
\end{figure*}

Figure \ref{fig:2d_map} illustrates the positions of particles in the disk tracked over 300 kyr of PSN evolution, with each particle assumed to have a density of 1 g/cm$^{-3}$. The particle trajectory shown in Fig. \ref{fig:2d_map} represents the mean trajectory of 500 simulated particles. The insets on the right display the trajectory of an individual particle. Unlike the mean trajectories, which remain confined near the midplane, the individual particle trajectories exhibit greater vertical variation. However, the average vertical trajectory remains close to $z$ = 0 AU, due to the disk's symmetry about the midplane. The top three panels depict the averaged trajectories of particles sized at 1 cm, 100 $\mu$m, and 1 $\mu$m, respectively, starting their migration from the midplane region of the disk, where the local temperature is 20 K (approximately 11 AU from the star). The bottom three panels exhibit the averaged trajectories of particles with the same size distribution, but originating from the midplane of the disk where the local temperature is 80 K (approximately 6.1 AU from the star). 

On average, particles initiating their trajectories in the 20 K midplane region (panels $a$, $b$, $c$) require over 1 Myr of PSN evolution to cross the methanol iceline. The vertical trajectory of a particle is highly dependent on particle size. Specifically, 1 cm particles remain more confined to the midplane and never reach heights greater than 0.63 AU, whereas 1 $\mu$m particles are more likely to attain higher elevations, reaching up to 1.53 AU (as shown in the insets at the right of Fig. \ref{fig:2d_map}). On average, particles ranging from 1 cm to 1 $\mu$m cross the 10.2--10.8 AU region after 300 kyr of PSN evolution.

On average, particles that begin their trajectory at a local temperature of 80 K in the PSN (see panels $d$, $e$, and $f$ in Fig. \ref{fig:2d_map}) require more than 300 kyr to cross the methanol iceline and sublimate: approximately 530 kyr for 1 cm particles, and over 1 Myr for 100 $\mu$m and 1 $\mu$m particles. After 300 kyr of PSN evolution, the mean radial trajectories of particles ranging from 100 $\mu$m to 1 $\mu$m intersect the $\sim$5.7 AU region, while the mean radial trajectory of 1 cm particles reaches the $\sim$4.4 AU region.
The vertical trajectories also exhibit a distribution based on particle size (as shown in the insets at the right of Fig. \ref{fig:2d_map}): 1 cm particles are confined to the midplane and never reach altitudes above approximately 0.53 AU, whereas 1 $\mu$m particles can attain higher altitudes, up to about 1.23 AU.

\begin{figure*}[h]
\centering
\includegraphics[width=\textwidth]{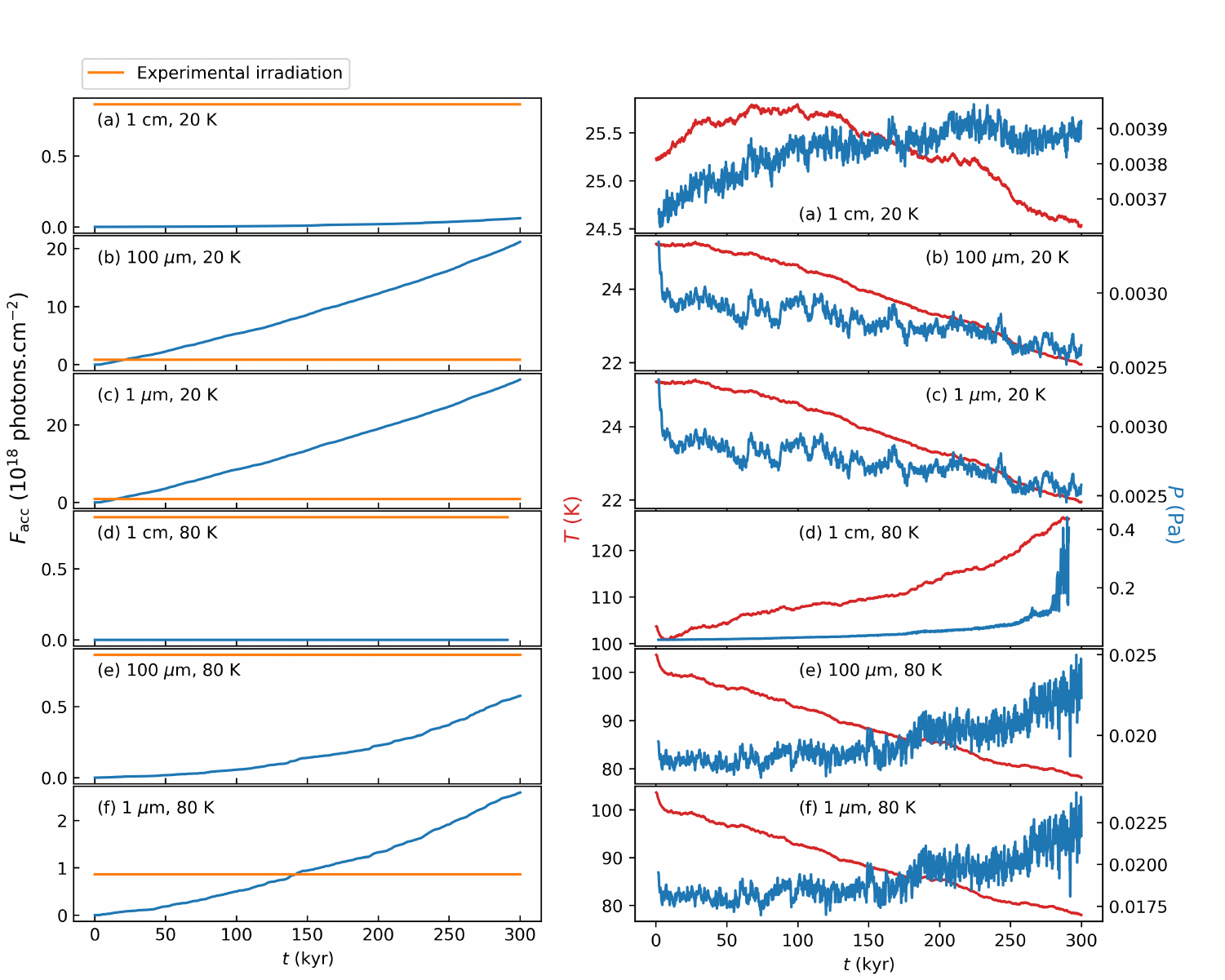}
\caption{Averaged accumulated irradiation, temperature, and pressure for particles sized 1 cm, 100 $\mu$m, and 1 $\mu$m, released at disk temperatures of 20 K (panels a, b, and c) and 80 K (panels d, e, and f). Left: Averaged accumulated irradiation of the particles over time (blue line) compared to the maximum irradiation accumulated by methanol ice during the experiments (orange line). Right: Averaged temperature (red line) and averaged pressure (blue line) experienced by the particles over time.}
\label{fig:variables}
\end{figure*}

Figure \ref{fig:variables} illustrates the average temperature and pressure of particles sized at 1 cm, 100 $\mu$m, and 1 $\mu$m along their trajectories in the PSN, as well as their average accumulated irradiation, assuming the same initial conditions as those displayed in Fig. \ref{fig:2d_map}. The average accumulated irradiation is determined by averaging the irradiation accumulated by each individual particle. In the following, terms such as accumulated irradiation, temperature, or pressure denote the average accumulated irradiation, average temperature, and average pressure of the 500 simulated particles.

On average, smaller particles measuring 1 and 100 $\mu$m, originating at a temperature of 20 K, receive the maximum accumulated irradiation in the experiments ($8.64 \times 10^{17}$ photons.cm$^{-2}$) after approximately 15.2--21.4 kyr of PSN evolution. In contrast, larger particles with a size of 1 cm would, on average, require approximately 911 kyr to accumulate the equivalent laboratory dose.

After 300 kyr of PSN evolution, smaller particles accumulate approximately 2.1--3.2 $\times 10^{19}$ photons.cm$^{-2}$, whereas larger pebbles accumulate about 6.2 $\times 10^{16}$ photons.cm$^{-2}$, on average. Additionally, at the end of this period, particles attain temperatures and pressures in the range of approximately 21--25 K and 2.5--4 mPa, on average, respectively.

Particles that start at a temperature of 80 K reach their maximum accumulated irradiation in the experiments after an average of 141 kyr and 359 kyr of PSN evolution for the 1 $\mu$m and 100 $\mu$m sized particles, respectively. In contrast, 1 cm sized particles do not reach the irradiation dose even after 1 Myr. By the end of their respective time spans, the 1 $\mu$m and 100 $\mu$m particles attain temperatures and pressures in the ranges of approximately 77--78 K and 21--23 mPa on average, respectively. In comparison, 1 cm sized particles reach temperatures and pressures of 127 K and 405 mPa on average, respectively.

% The 1 cm-sized particles reach the irradiation threshold in 201.5 kyr, while the 100 $\mu$m-sized particles reach the irradiation threshold in 194.37 kyr, and the 1 $\mu$m-sized particles reach the irradiation threshold in 174.9 kyr. In 300 kyr, the 1 cm-sized particles reach an average accumulated irradiation of $2.53 \cdot 10^{18}$ photons.cm$^{-2}$, an average temperature of 642 K, and an average pressure of 87.8Pa. In 300 kyr, the 100 $\mu$m-sized particles reach an average accumulated irradiation of $2.36 \cdot 10^{18}$ photons.cm$^{-2}$, an average temperature of 626 K and an average pressure of 85.1 Pa. In 300 kyr, the 1 $\mu$m-sized particles reach $2.87 \cdot 10^{18}$ photons.cm$^{-2}$, an average temperature of 624 K and an average pressure of 84.3 Pa. In average, the conditions where the 1 cm-sized particles cross the iceline is 296 K and 56.3 Pa, while for the 100 $\mu$m-sized particles it is 339 K and 66.2 Pa, and for the 1 $\mu$m-sized particles it is 377 K and 78.4 Pa
        
\section{Discussion and conclusion} 
\label{sec:discussion}

We investigated the influence of disk parameters on the mean trajectory and the average accumulated irradiation of the particles.

The opacity used in this model is the mean Rosseland opacity, derived from \citet{Bel1994}, which is averaged across all frequencies and may inadequately represent UV-range photon irradiation. To account for potential changes in opacity, we employed an opacity of 48 cm$^2$/g following the model of \citet{Bir2018} for particles up to 10 cm in size, averaged over the UV wavelength range of 122--160 nm used in the experiments by \citet{Ten2022}. We conducted new simulations for 1 $\mu$m particles, which are more likely to receive higher irradiation doses. 

After 300 kyr of PSN evolution, the 1 $\mu$m particles released at 11 AU ($\sim$20 K) receive an average accumulated irradiation dose of about $\sim$3.2$\times$10$^{16}$ photons/cm$^{-2}$ with this new opacity, which is several orders of magnitude lower than the average accumulated irradiation dose of $\sim$3.2$\times$10$^{19}$ photons/cm$^{-2}$ in our nominal case. Similarly, 1 $\mu$m particles released at 6.1 AU ($\sim$80 K) receive an average accumulated irradiation dose of about $\sim$2.3$\times$10$^{16}$ photons/cm$^{-2}$ with the new opacity, compared to the $\sim$2.6$\times$ 10$^{18}$ photons/cm$^{-2}$ in our nominal case. This new opacity of 48 cm$^2$/g results in a lower irradiation of the particles in our model, and on average, the particles do not receive the experimental irradiation dose. However, it should be noted that although this opacity is more suitable for UV radiation, it remains constant throughout the disk and does not vary with temperature or density, unlike our nominal opacity. Due to this significant limitation, we used the standard opacity data from \cite{Bel1994}.

An increase in the viscosity parameter, $\alpha$, generally extends the particle trajectory in the PSN and tends to increase their irradiation exposure. For instance, with $\alpha = 1 \times 10^{-2}$ (10 times larger than the nominal value), the mean trajectory of 500 particles sized at 1 cm and released at 20 K begins at approximately 14 AU, which is about 3 AU farther than in our nominal computation. The particles drift inward in the PSN at a velocity approximately 3.5 times higher. They do not cross the methanol iceline even after 1 Myr of PSN evolution. Furthermore, the irradiation accumulated by the particles matches the laboratory dose after only approximately 16 kyr of PSN evolution, more than 50 times faster than in our nominal case. This acceleration happens because the outer region of the disk is optically thinner than the inner region, and a more viscous disk allows particles to reach greater altitudes.
Particles measuring 1 cm and released at a local temperature of 80 K commence their journey at approximately 6.6 AU (slightly farther than in our nominal case) and drift approximately 4 times faster than in our nominal case. However, they do not cross the methanol iceline, even after 1 Myr of PSN evolution. Furthermore, they still do not receive the experimental irradiation dose after this duration. On the other hand, 1 $\mu$m sized particles released at a local temperature of 80 K receive the laboratory irradiation dose in approximately 36 kyr of PSN evolution (roughly 4 times faster than in our nominal case) and drift approximately 5 times faster than in our nominal case. However, they still do not cross the methanol iceline during the first million years of PSN evolution.

Conversely, assuming $\alpha = 1 \times 10^{-4}$ (10 times less than the nominal value), 1 cm particles released at a local temperature of 20 K begin their migration at a distance of approximately 7.3 AU (about 3.7 AU closer to the star than in our nominal case) and drift inward in the PSN at a velocity approximately 3 times lower. They do not reach the irradiation threshold (accumulating approximately 10$^{11}$ times less irradiation than in our nominal case) and do not cross the methanol iceline, even after 1 Myr of PSN evolution.
Similarly, particles measuring 1 cm released at a local temperature of 80 K start their drift at a radius of approximately 4.6 AU, very close to the methanol iceline, which is located at around 3.9 AU. They drift inward approximately 3 times more slowly than in our nominal case and do not cross the methanol iceline within 1 Myr of PSN evolution. Additionally, they do not receive the irradiation dose on this timescale.
Particles measuring 1 $\mu$m released at a local temperature of 80 K also start at 4.6 AU and drift approximately 7 times more slowly than in our nominal case. They also fail to cross the iceline within 1 Myr of PSN evolution and experience irradiation at a rate approximately 7 times lower than in our nominal case. Therefore, the particles also fail to receive the irradiation dose during this simulation time frame.

Another key parameter is the initial disk's mass accretion rate. A variation by a factor of 2 does not lead to significant differences. For instance, 1 cm sized particles starting from a local temperature of 20 K evolving in a disk with $\dot{M}_0$ = $5 \times 10^{-7}$ M$_\sun$/year (2 times higher than in our nominal case) receive approximately 1.6 times less irradiation and drift around 1.1 times faster. With $\dot{M}_0$ = $1 \times 10^{-7}$ M$_\sun$/year (2.5 times less than our nominal value), those particles receive roughly 1.5 times more irradiation and drift approximately 1.1 times more slowly.
The 1 cm sized particles released at a local temperature of 80 K with $\dot{M}_0$ = $5 \times 10^{-7}$ M$_\sun$/year accumulate 10$^{14}$ more irradiation than in our nominal case, and drift $\sim$1.7 times more slowly. However, with $\dot{M}_0$ = $1 \times 10^{-7}$ M$_\sun$/year, those particles accumulate 10$^{12}$ times more irradiation than in our nominal case and drift 1.9 times more slowly. 1 $\mu$m sized particles released at a local temperature of 80 K with $\dot{M}_0$ = $5 \times 10^{-7}$ M$_\sun$/year drift inward approximately 1.1 times faster than in our nominal case and receive the irradiation dose about 120 kyr later, at 261 kyr of PSN evolution. With $\dot{M}_0$ = $1 \times 10^{-7}$ M$_\sun$/year, those particles drift approximately 1.2 times more slowly than in our nominal case and receive the irradiation dose after 144 kyr of PSN evolution, similar to our nominal case. This suggests that reducing $\dot{M}_0$ cools the disk, leading to lower radial velocities for particles; however, even an initial mass accretion rate that is half of the nominal value does not significantly alter the results.

Another parameter influencing the trajectory of particles is their density. 1 cm sized particles at 20 K with twice the nominal density (2 instead of 1 g.cm$^{-3}$) receive approximately 20 times less irradiation ($2.2\times10^{14}$ instead of $4.4\times10^{15}$ photons.cm$^{-2}$ in 100 kyr of PSN evolution) because they remain more confined in the midplane, and they drift 1.6 times faster. Conversely, with a density half that of our nominal density (0.5 instead of 1 g.cm$^{-3}$), the particles are likely to reach higher altitudes, experience nearly 10 times more irradiation ($4.3\times10^{16}$ photons.cm$^{-2}$ over 100 kyr of PSN evolution), and drift 1.5 times more slowly. In contrast, 1 $\mu$m sized particles released at 20 K with a density of 0.5 g.cm$^{-3}$ exhibit no significant differences from our nominal case in terms of trajectories or irradiation (the less dense particles accumulate $8.5\times10^{18}$ instead of $8.44\times10^{18}$ photons.cm$^{-2}$ after 100 kyr of PSN evolution). Similar conclusions can be drawn with particles twice as dense, which are irradiated with $8.35\times10^{18}$ photons.cm$^{-2}$ after 100 kyr of PSN evolution.
Finally, 1 cm particles released at a local temperature of 80 K show similar results when we vary their density: Particles with a density twice as high as our nominal case (2 g cm$^{-3}$ instead of 1 g cm\(^{-3}\)) drift 1.7 times faster and receive less irradiation (less than 1 instead of $1.1\times10^{6}$ photons cm$^{-2}$ over 100 kyr of PSN evolution). Conversely, particles with a density lower than in our nominal case (0.5 g cm$^{-3}$ instead of 1 g cm$^{-3}$) drift 1.5 times more slowly and receive more irradiation ($3.3\times10^{9}$ photons cm$^{-2}$ over 100 kyr of PSN evolution). The variation in particle density affects their radial evolution only slightly but significantly impacts their vertical trajectories and, therefore, their irradiation, particularly for larger particles measuring 1 cm.

%Particles starting from a local temperature of 20 K end up at $\sim$10.5 AU after 150 kyr of PPD evolution (similar than our nominal case). They reach the accumulated irradiation threshold after 13.1 kyr of PPD evolution (instead of 13.5 kyr in our nominal case), but do not cross the iceline in 1 Myr of PPD evolution. Particles of 0.5 g.cm$^{-3}$ from 80 K cross the iceline in $\sim$109 kyr (20 kyr earlier than our nominal case), and they are accreted to 4.7 AU in 135.6 kyr (similar to our nominal case). They would reach the maximum accumulated irradiation in 254 kyr, which is 60 kyr after our nominal case. 1 cm-sized particles do not show significant differences when there is a density variation by a factor of 2. When they start where t= 20K, and with a density of 2 g.cm-3, they reach the irradiation dose after 22.341 kyr, while with a density of 0.5 g.cm-3 they reach it after 23.372 kyr, so 1 kyr later, instead of 22.684 kyr in our nominala case. Starting from 80 K, centimetric particles with a density of 2 g.cm-3 end up at a location of 5.52 AU after 58 kyr, while with a density of 0.5 g.cm-3, they end up at 5.55 AU, instead of 5.53 AU in our nominal case. To sum up, a variation of the particles density by a factor of 2 do not vary significantly the results.

Our findings indicate that 1 cm particles experience lower levels of irradiation due to gravitational forces anchoring them in the midplane. Conversely, smaller particles measuring 100 and 1 $\mu$m are prone to ascend to higher altitudes and thus experience increased irradiation. When released at a local temperature of 80 K (approximately 6.1 AU from the star), these particles require 359 and 141 kyr of PSN evolution, respectively, to accumulate sufficient irradiation for substantial molecular diversity to emerge \citep{Ten2022}. However, the irradiation timescale drastically reduces to less than 22 kyr for the formation of COMs when particles are released at a local temperature of 20 K (around 11 AU from the star). This notably brief timescale, relative to disk evolution, suggests that methanol-rich ices transported by small particles in the outer disk regions (particularly those beyond approximately 11 AU) efficiently convert into COMs before consolidating into larger solids. The growth timescale of particles is approximately 2.5 kyr at 11 AU after 10 kyr of PSN evolution, is shorter at greater distances, and is longer closer to the star (approximately 3.5 kyr at 6.1 AU). This growth timescale increases over time, reaching 80 kyr at 11 AU after 1 Myr of PSN evolution \citep{Bir2012,Agu2020}. 

Consequently, COMs could be prevalent in the outer regions of the PSN. Furthermore, particles smaller than 1 cm will not cross the iceline and therefore will not release molecules in the vapor phase during the first million years of PSN evolution if they are released at distances greater than $\sim$6 AU. To cross the methanol iceline, particles must be either similar to or larger than the 1 cm particles simulated here or be released closer to the iceline. Additionally, \cite{Mis2019} has demonstrated that the growth and fragmentation of particles strongly influence their trajectories. Integrating the evolution of grain growth with their trajectories, and consequently their irradiation, presents an interesting avenue for future research.

Furthermore, it is crucial to consider that the experimental photon flux exceeds that in our simulations. In the experiments, an irradiation of 10$^{13}$ photons s$^{-1}$ cm$^{-2}$ results in an experimental ratio of 7.2 photons per molecule, compared to a maximum irradiation in the disk of 10$^{8}$ photons s$^{-1}$ cm$^{-2}$ in our simulations. The higher photon dose rate in laboratory experiments, relative to PSN conditions, likely enhances reactivity within the sample. This increased reactivity can promote the formation of more diverse and complex organic products, as recently formed radicals are exposed to additional photons. Conversely, the lower dose rate typical in PSN conditions allows for the recombination of radicals into their precursors. Hence, the same total irradiation dose would likely lead to the formation of fewer COMs in the PSN compared to laboratory experiments.

Rocky particles coated with an organic mantle (OMGs) exhibit greater adhesiveness compared to particles composed solely of organics or rocks \citep{Kou2002,Hom2019,Bis2020}. The enhanced fragmentation threshold velocity of OMGs within the 1.1–-3.4 AU region (corresponding to a local temperature range of approximately 200-400 K) facilitates the formation of planetesimals \citep{Kou2002,Hom2019}. Nevertheless, our simulations suggest that particles released at local temperatures of 20 K or 80 K typically never reach this zone, even after 1 Myr of PSN evolution. Future studies should prioritize examining the evolution of newly formed COMs to explore their potential condensation as mantles on rocky particles within the region delineated by \cite{Kou2002}. Another intriguing avenue of research would involve investigating the formation of COMs from molecules other than methanol, which may sublime at higher temperatures or require greater irradiation to develop such molecular diversity. To investigate this, integrating a chemical photoreaction model, such as those proposed by \cite{Tak2022} or \cite{Och2024}, with the particle transport model could be highly beneficial. This approach could potentially facilitate the development of organic mantles around pebbles in regions characterized by the stickiest conditions in simulations.

Our primary focus has been on interstellar cosmic irradiation, which has led us to overlook other significant processes such as cosmic ray interactions with hydrogen gas. These interactions generate secondary UV radiation, albeit with low irradiation values of $1.4 \times 10^3$ photons cm$^{-2}$ s$^{-1}$, in contrast to our nominal values \citep{Pra1983}. Additionally, the effect of electron irradiation, though often disregarded, remains an important consideration. Furthermore, potential hydrogenation processes, known for their ability to destruct molecules \citep{Lin2011}, have not been factored into our analysis. These processes, despite appearing negligible, can contribute to irradiation in opaque regions of the disk, such as the midplane. Future studies should thus attempt to accurately quantify their impacts.

Planets, moons, and asteroids have grown from material originating from the PSN. Thus, our model offers vital insights into the drift ranges and thermodynamic conditions under which pebbles can foster molecular diversity from methanol ice. Should the distance range for COMs formation, as delineated by our model, coincide with the formation zones of planets and other celestial bodies, these bodies might have, at least partially, evolved from these COMs. Given the analogous properties between PPDs and circumplanetary disks, the particle transport module utilized in our study holds promise for application in circumplanetary disk models, thereby shedding light on the role of COMs in moon formation.
 
%Experimental data showed that the molecular diversity can be achieved with enough UV irradiation accumulated on methanol ice, which can stick on rocky pebbles in the PPD. And considering our modeling, those particles can easily reach this dose to form COMs, depending on their starting positions and their sizes. Particles located closer than 6 AU from the star release methanol into their vapour phase too early, while particles starting their trajectory further from 11 AU reach sufficient irradiation more quickly, but are less likely to release their molecular diversity into the vapour phase. Particles released in the $\sim$6-11 AU range are more likely to form molecular diversity, then release it in the vapor phase as they cross the iceline. Moreover, micrometric particles can accumulate more irradiation than centrimetric particles. 
%Future experiments will have to consider important variations of irradiation of the icy samples corresponding to the upward and downward trajectories of particles in the disk. It would be interesting to also explore other configurations relevant to planetary formation, such as more important irradiation doses applied to icy samples at very low temperatures. This would correspond to the evolution of particles in the formation region of the outer planets far from the star. Another perspective, for both modeling and experiment, could be to take into account other interactions such as hydrogenation or interaction with electron flux.

\begin{acknowledgements}
O. Mousis acknowledges support from CNES and the Programme National de Plan\'etologie (PNP) of CNRS/INSU. The project leading to this publication has received funding from the Excellence Initiative of Aix-Marseille Universit\'e--A*Midex, a French ``Investissements d’Avenir program'' AMX-21-IET-018. This research holds as part of the project FACOM (ANR-22-CE49-0005-01\_ACT) and has benefited from a funding provided by l'Agence Nationale de la Recherche (ANR) under the Generic Call for Proposals 2022. This material is based upon work supported by NASA'S Interdisciplinary Consortia for Astrobiology Research (NNH19ZDA001N-ICAR) under award number 19-ICAR19\_2-0041. AB acknowledges support from CNES. This work was supported by the Programme National de Planétologie (PNP) of CNRS-INSU cofunded by CNES. A.B. acknowledges support from the French government under the France 2030 investment plan, as part of the Initiative d'Excellence d'Aix-Marseille Université—A*MIDEX AMX-21-PEP-032. We would like to thank the anonymous referees for their helpful comments, which significantly improved the quality of this paper.
\end{acknowledgements}

\begin{appendix}

\section{Opacity values} \label{appendix:opacity}

The mean Rosseland opacity, denoted as $\kappa_\textup{R}$, is a frequency-averaged quantity that depends on the gas density $\rho_\mathrm{m}$ and the temperature $T_m$ in the midplane. It is expressed as $\kappa_\textup{R} = \kappa_0 \rho_\mathrm{m}^a T_m^b$, where $\kappa_0$, $a$, and $b$ are parameters determined through observations by \cite{Bel1994}. Detailed values for these parameters are provided in Table \ref{tab:opacities}.

\begin{table}[h!]
\centering
\caption{Opacity values adopted for different regions, sourced from \cite{Bel1994}.}
\begin{tabular}{lccc}
\hline
\hline
\smallskip
Regime & $\kappa_0$ [cm$^{2}$/g] & $a$ & $b$ \\
\hline
Ice grains  & $2\times10^{-4}$  & 0  & 2 \\
Ice grains evaporation   & $2\times10^{16}$ & 0 & -7  \\
Dust grains  & 0.1 & 0 & 1/2 \\
Dust grains evaporation & $2\times10^{81}$ & 1 & -24  \\
Molecules  & $10^{-8}$ & 2/3 & 3 \\
H-scattering & $10^{-36}$ & 1/3 & 10 \\
Bound-free and free-free & $1.5\times10^{20}$ & 1 & -5/2 \\
Electron scattering & 0.348 & 0 & 0 \\
\hline
\end{tabular}
\label{tab:opacities}
\end{table}

\section{Stokes number} \label{appendix:regimes}

\begin{figure}[h!]
\centering
\includegraphics[width=\linewidth]{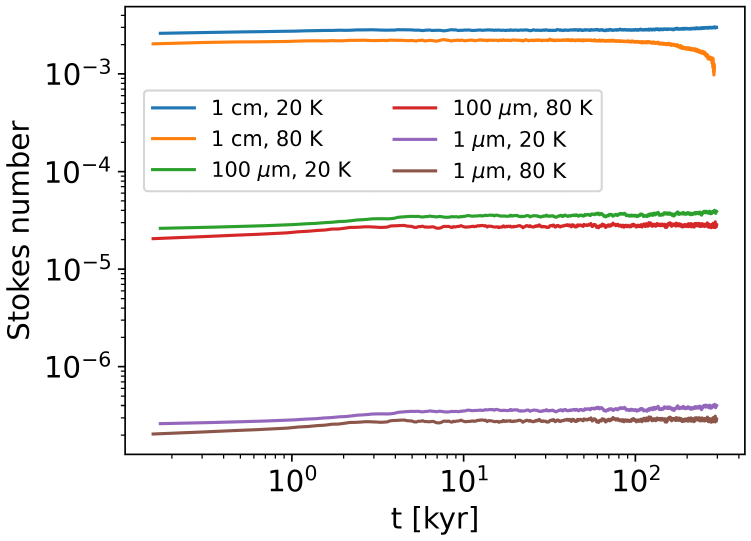}
\caption{Stokes number computed for 500 simulated particles in each size category (1 cm, 100 $\mu$m, and 1 $\mu$m) and each based on the average temperature and density of the disk they encounter and their mean trajectories. Particles were released at temperatures of 20 K and 80 K.}
\label{fig:st_part}
\end{figure}

\begin{figure}[h!]
\centering
\includegraphics[width=\linewidth]{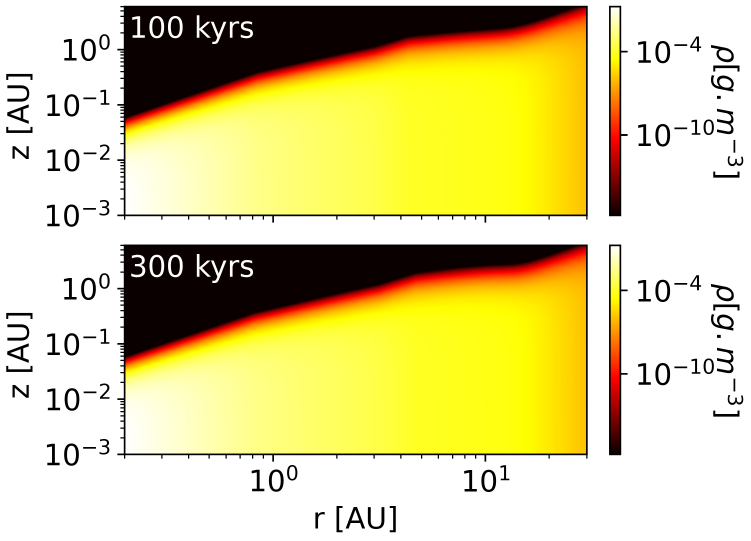}
\caption{Two-dimensional density profile of the disk after 100 and 300 kyr of evolution.}
    \label{fig:rho_2d}
\end{figure}

\begin{figure}[h!]
\centering
\includegraphics[width=\linewidth]{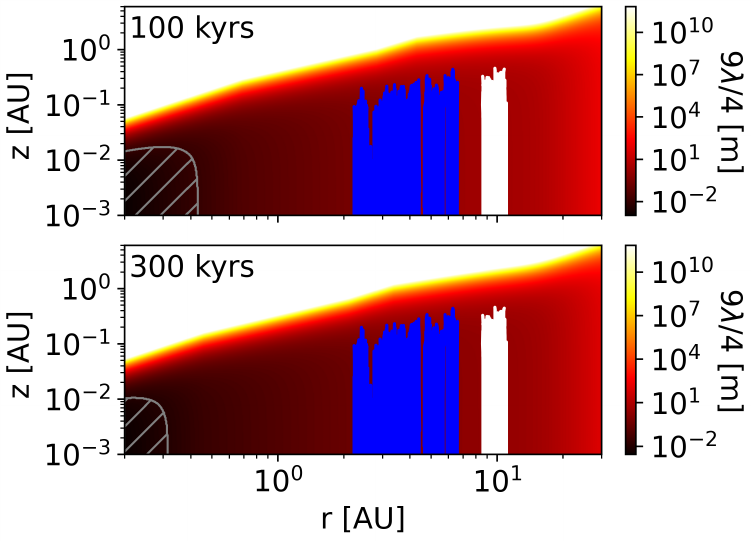}
\caption{Mean free path of the gas, $\lambda$, multiplied by 2.25 after 100 and 300 kyr of PSN evolution. Particles smaller than 9$\lambda$/4 are in the Epstein regime, while larger particles fall into the Stokes regime. Blue and white lines represent the median trajectories of 500 simulated 1 cm particles released at 80 K and 20 K, respectively. The gray hatching highlights the region where 1 cm and smaller particles transition from the Epstein regime to the Stokes regime.} 
\label{fig:epstein_2d}
\end{figure}

In the PPD, particles can either be coupled to the gas or experience radial drift due to decoupling from the gas. The Stokes number governs the radial velocity component through Equation \ref{eq:v_r}. A lower Stokes number indicates stronger coupling to the gas, while the influence of radial drift peaks when the Stokes number is close to 1 \citep{Jos2014,Dra2023}.
The Stokes number is defined as St=t$_s$$\Omega_K$ (with $t_s=\frac{\rho_s R_s}{\rho_g \sqrt{8/\pi} c_s} $, $c_s = \sqrt{R_g T_m / \mu}$ and $\Omega_K = \sqrt{G M_{\star} / r^3}$). Thus, it depends on the disk density, temperature, and the particles' position and size. Specifically, larger particles have a higher Stokes number (as shown in Fig. \ref{fig:st_part}), and denser disks result in a lower Stokes number (with the density profile displayed in Fig. \ref{fig:rho_2d}). Moreover, the stopping time varies depending on whether the particles are in the Stokes or Epstein regime \citep{Jos2014,Dra2023}. Particles with a size smaller than 9$\lambda$/4 are in the Epstein regime, as described in Sect. \ref{sec:transport}. In our nominal cases, we are always in the Epstein regime (cf. Fig. \ref{fig:epstein_2d}, which shows the 9$\lambda$/4 map of the disk).

\begin{figure}[h!]
    \centering
    \includegraphics[width=\linewidth]{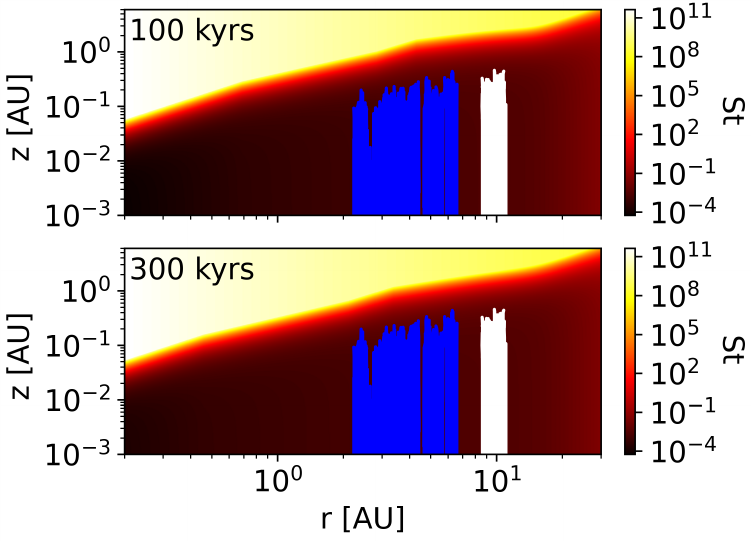}
    \caption{Stokes number in the disk after 100 and 300 kyr of PSN evolution, calculated for 1 cm sized particles. The Stokes number is lower for smaller particles. As in Fig. \ref{fig:epstein_2d}, the median trajectories of 500 simulated 1 cm particles, released at 80 K and 20 K, are represented by blue and white lines, respectively.}
    \label{fig:stokes_2d}
\end{figure}

Figure \ref{fig:stokes_2d} shows the Stokes number for 1 cm sized particles in the PPD, along with the median trajectory of these particles. Figure \ref{fig:st_part} presents the Stokes number for particles of various sizes, calculated using the mean density and mean temperature encountered by the particles and their mean trajectories, in each of our nominal cases. It is evident that particle size is the primary factor influencing the Stokes number. Specifically, 1 cm sized particles have a Stokes number ranging from \(1 \times 10^{-3}\) to \(3 \times 10^{-3}\), 100 $\mu$m sized particles range from 1 $\times 10^{-5}$ to 3 $\times 10^{-5}$, and 1 $\mu$m sized particles range from 1 $\times 10^{-7}$ to 4 $\times 10^{-7}$.

\end{appendix}

\end{document}